\documentclass[10pt,twocolumn,letterpaper]{article}

\usepackage[pagenumbers]{iccv} 

\newcommand{\ourmodel}{MAVFlow\xspace}

\definecolor{iccvblue}{rgb}{0.21,0.49,0.74}
\definecolor{iccvviolet}{rgb}{0.54, 0.17, 0.89}
\usepackage[pagebackref,breaklinks,colorlinks,allcolors=iccvblue]{hyperref}
\usepackage[accsupp]{axessibility}  

\usepackage{pifont} 
\usepackage{multirow}
\usepackage{xcolor}
\usepackage{kotex}
\usepackage{footnote}
\usepackage{booktabs}
\usepackage{subcaption}
\usepackage{afterpage}
\newcommand{\cmark}{\textcolor{PineGreen}{\ding{51}}}
\newcommand{\xmark}{\textcolor{Red}{\ding{55}}}

\title{\ourmodel: Preserving Paralinguistic Elements with Conditional Flow Matching for Zero-Shot AV2AV Multilingual Translation}

\author{
    Sungwoo Cho$^{1}$ \quad 
    Jeongsoo Choi$^{2}$ \quad 
    Sungnyun Kim$^{1}$ \quad 
    Se-Young Yun$^{1}$
    \vspace{3pt}\\
    {$^1$KAIST AI \quad $^2$KAIST EE} \\
    {\tt\small \{peter8526, jeongsoo.choi, ksn4397, yunseyoung\}@kaist.ac.kr}
}

\begin{document}
\maketitle
\begin{abstract}

Despite recent advances in text-to-speech (TTS) models, audio-visual-to-audio-visual (AV2AV) translation still faces a critical challenge: maintaining speaker consistency between the original and translated vocal and facial features. To address this issue, we propose a conditional flow matching (CFM) zero-shot audio-visual renderer that utilizes strong dual guidance from both audio and visual modalities. By leveraging multimodal guidance with CFM, our model robustly preserves speaker-specific characteristics and enhances zero-shot AV2AV translation abilities. For the audio modality, we enhance the CFM process by integrating robust speaker embeddings with x-vectors, which serve to bolster speaker consistency. Additionally, we convey emotional nuances to the face rendering module. The guidance provided by both audio and visual cues remains independent of semantic or linguistic content, allowing our renderer to effectively handle zero-shot translation tasks for monolingual speakers in different languages. We empirically demonstrate that the inclusion of high-quality mel-spectrograms conditioned on facial information not only enhances the quality of the synthesized speech but also positively influences facial generation, leading to overall performance improvements in LSE and FID score. Our code is available at \href{https://github.com/Peter-SungwooCho/MAVFlow}{https://github.com/Peter-SungwooCho/MAVFlow}.

\end{abstract}
\vspace{-0.3cm}
\section{Introduction}
\label{sec:intro}
With the rapid proliferation of multimedia content and increasing cross-cultural interactions, the expansion from one language to another has become essential to enrich user engagement and comprehension. Traditional approaches in language translation, such as subtitle processing via neural machine translation (NMT)~\cite{matusov2019customizing} or single-modality methods like speech-to-speech translation and dubbing~\cite{federico2020speech}, often fail to deliver a fully immersive experience. For instance, in dubbed films, discrepancies between the original visual content and the dubbed audio can lead to unnatural lip synchronization and a mismatch between the expected and dubbed voices. Such inconsistencies disrupt the viewer's concentration and diminish the overall experience. The adverse effects of audio-visual incongruence on user perception have been substantiated by the McGurk effect \cite{mcgurk1976hearing}; notably, when the dubbed voice deviates from the expected voice of original actors, the naturalness of the content significantly deteriorates \cite{koolstra2002pros}.

\begin{figure}[!t]
    \centering
    \includegraphics[width=\linewidth]{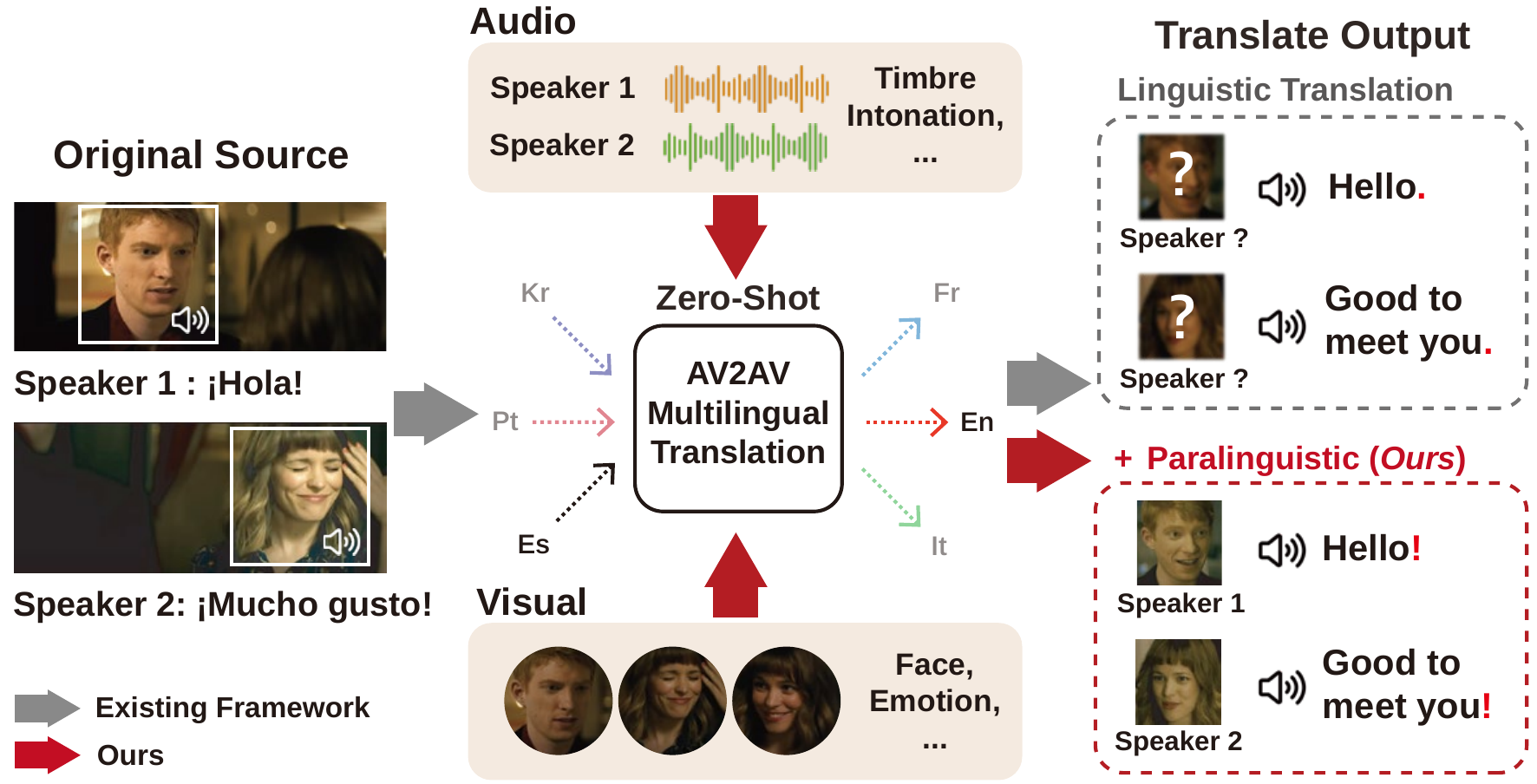}
    \caption{
    Overview of the existing audio-visual translation (AV2AV) framework. Conventional AV2AV methods primarily focus on linguistic content, often neglecting crucial paralinguistic features, such as speaker identity and emotional nuance, which are essential for maintaining consistent speaker characteristics.}
    \vspace{-0.3cm}
    \label{fig:figure1}
\end{figure}

At a fundamental level, the transition from single-modality to dual-modality translation is achievable via cascaded approaches. A typical pipeline involves using an automatic speech recognition\,(ASR) model\,\cite{anwar2023muavic, radford2023robust} to transcribe the source audio into text, subsequently applying NMT~\cite{fan2021beyond, costa2022no} for language conversion, and finally synthesizing speech via text-to-speech\,(TTS) systems\,\cite{casanova2022yourtts, wang2023neural, casanova2024xtts} in conjunction with talking face generation\,(TFG) models\,\cite{zhou2019talking, prajwal2020lip, park2022synctalkface}. However, such cascaded methods are complex and often suffer from significant information loss due to repeated modality transformations and intermediate text representations. To overcome these challenges, direct audio-visual-to-audio-visual (AV2AV) translation approaches have been introduced \cite{huang2023av, choi2024av2av, cheng2023transface}. These methods bypass textual representations by leveraging discrete units obtained from self-supervised multimodal models (\eg, multilingual AV-HuBERT~\cite{shi2022learning, choi2024av2av}), enabling a more direct and efficient translation of audio-visual source inputs.

Despite these advances, the AV2AV translation still faces a critical challenge: preserving speaker consistency (\eg, tone, pitch, or facial expressions) between the original and translated audio-visual data (as shown in Figure~\ref{fig:figure1}). This limitation is exacerbated by the absence of datasets wherein the same speaker articulates identical content in multiple languages, necessitating zero-shot strategies for speaker preservation. Current AV2AV approaches~\cite{choi2024av2av, cheng2023transface} adopt a simple architecture that relies on using speaker embeddings, combining speaker-specific d-vectors~\cite{variani2014deep} within the speech vocoder~\cite{kong2020hifi, prajwal2020lip}. This yet underexplored structure of AV2AV, which does not employ advanced conditional generation techniques, still restricts its ability to maintain speaker consistency.
Moreover, generating audio and visual components are separated, using a single-modality embedding in each part. This has limitations from a multimodal perspective since only the reference audio is used, while visual cues could also be considered for speech generation.

In this study, to \textit{preserve paralinguistic elements} of speakers during the \textit{linguistic translation}, we propose \ourmodel, a Conditional Flow Matching (CFM)~\cite{mehta2024matcha} based zero-shot audio-visual renderer that leverages dual guidance from audio and visual modalities. Notably, for ideal multilingual translation scenarios, a speaker’s voice characteristics and facial information (\eg, appearance or emotion) must remain consistent regardless of language~\cite{yang2020large, brannon2023dubbing, adhikary2023travid}. Based on this hypothesis, we adopt a guidance strategy that utilizes speaker embeddings from audio and emotion embeddings from visual input. This strategy enables the complementary capture of paralinguistic information, which are commonly shared across both audio and visual modalities.

Furthermore, \ourmodel leverages the Optimal Transport\,(OT) CFM's structural advantages in integrating a multimodal guidance. OT-CFM\,\cite{mehta2024matcha} facilitates learning a conditional speech distribution, enhancing zero-shot performances and guidance-based control, making it an ideal approach for preserving paralinguistics in AV2AV translation.
\ourmodel incorporates x-vector-based speaker embeddings~\cite{snyder2018x} of audio and facial emotion embeddings~\cite{toisoul2021estimation} of visual inputs, directly guiding the flow matching generative module. Additionally, OT-CFM significantly improves speech synthesis performance and enables more efficient sampling with fewer steps. Thus, \ourmodel achieves enhanced speaker consistency while producing seamless audio-visual translation results in cross-lingual scenarios. Our contributions are summarized as follows:

\begin{itemize}[leftmargin=10pt, label={$\circ$}]
    \item We propose \ourmodel, integrating discrete speech units with OT-CFM to efficiently synthesize high-quality mel-spectrograms for advanced audio-visual translation.
    \item We transmit paralinguistic speaker characteristics from both audio and visual modalities within the latent space of the OT-CFM model, thereby achieving robust zero-shot capabilities in cross-lingual scenarios.
    \item  We empirically demonstrate that our dual guidance improves the consistency of speaker identity in synthesized speech by an average of 36\% on the MuAViC dataset\,\cite{anwar2023muavic}, while enhancing face generation with gains in lip-sync accuracy ($+$0.87) and visual quality score ($-$0.61) on textless system.
    \item  We also confirm that \ourmodel effectively represents emotion in both audio and visual generation on the CREMA-D dataset\,\cite{cao2014crema}.
\end{itemize}

\section{Related Works}
\label{sec:related_works}
\subsection{Spoken Language Translation}
\label{ssec:spoken_lang_translation}
Spoken Language Translation (SLT) aims to convert spoken language in one language into another language, promoting natural cross-lingual interaction. Traditional SLT typically adopt cascaded approaches~\cite{ney1999speech,matusov2006integration} for speech-to-speech translation (S2ST), chaining ASR, NMT, and TTS. Although widely used, cascaded methods suffer from cumulative errors, latency, and loss of speaker-specific prosodic and paralinguistic features~\cite{weiss2017sequence,jia2019direct}. To reduce these issues, research has shifted toward end-to-end SLT methods that translate speech directly~\cite{jia2022translatotron,nachmani2024translatotron}, and even textless approaches that eliminate the reliance on textual representations~\cite{lee2021textless,kim2024textless}.

Despite advances, S2ST systems primarily focus on audio signals, often neglecting the alignment between translated speech and visual information, which is crucial for cross-lingual scenarios such as video conferencing or dubbing. This can lead to lip-sync mismatches~\cite{mcgurk1976hearing, koolstra2002pros}, disrupting realistic multimodal experiences. Speech-driven TFG has been explored to synchronize video with translated speech~\cite{kr2019towards, waibel2023face}. More recently, TransFace~\cite{cheng2023transface} and AV2AV~\cite{choi2024av2av} jointly generate synchronized audio and visual outputs. However, achieving natural cross-lingual experiences remains challenging, particularly in preserving speaker identity and maintaining emotional consistency.

\subsection{Flow Matching for Generative Modeling}
\label{ssec:conditional_flow_matching} 
Generative models based on diffusion process~\cite{ho2020denoising, song2020denoising} have demonstrated remarkable performance across various domains, including image~\cite{dhariwal2021diffusion, rombach2022high}, speech~\cite{popov2021grad, jeong2021diff}, audio~\cite{kong2020diffwave, chen2021wavegrad2}, and video~\cite{ho2022video, blattmann2023stable}, by iteratively denoising to generate high-fidelity outputs. Despite their impressive quality, diffusion-based models often require numerous sampling steps, limiting their practicality for real-time or large-scale applications. Flow matching~\cite{lipman2022flow, liu2022flow} models address this limitation by learning direct stochastic paths between distributions, enabling efficient and high-quality generation in fewer steps~\cite{liu2023instaflow}. This approach has been successfully applied to various tasks~\cite{le2023voicebox, mehta2024matcha, esser2024scaling, ma2024sit}.

Recent works have explored conditioning flow matching models to enable following conditions while maintaining high-quality generation, becoming prominent in generative modeling. Additionally, conditioning with multiple inputs has emerged as a promising direction~\cite{mou2024t2i}, and conditional flow matching model successfully leverage these conditions to produce aligned and controllable outputs~\cite{vyas2023audiobox, gao2024lumina}. However, effectively extracting and utilizing relevant information from multimodal signals for conditioning remains in its early stages.

\section{Preliminaries}
\label{sec:preliminaries}

\subsection{Audio-Visual Speech Unit Translation}
\label{subsec:unit_translation}
Recent advances in direct audio-visual-to-audio-visual translation have leveraged discrete speech units to bypass intermediate text transcription, thereby avoiding delays and error propagation of cascaded systems and expanding applicability~\cite{cheng2023transface}. To obtain translated AV units, our system follows two-stage procedure. First, we extract discrete AV units from an input sequence using the unit extractor, m-AVHuBERT~\cite{choi2024av2av}, which has been pretrained on 7,000 hours of multilingual audio-visual data. Second, we pass the extracted discrete units to a unit-to-unit\,(U2U) translation module~\cite{choi2024av2av}, which translates them into the counterparts of target language. The translated units are subsequently converted into intermediate features (mel-spectrograms) via CFM, and finally transformed back into audio-visual form through the vocoder and face decoder. Notably, both the unit extractor and the U2U translation module are identical to those used in our prior work~\cite{choi2024av2av}, thus ensuring consistency in performance.

\subsection{Optimal Transport CFM}
\label{subsec:conditional_flow_matching}
Conditional Flow Matching (CFM) is a framework that leverages conditional flows to train generative models, particularly applied to generate mel-spectrograms in audio synthesis tasks~\cite{mehta2024matcha, du2024cosyvoice}. Unlike conventional flow-based models, which learn a bijective mapping between a simple prior distribution (\eg, Gaussian noise) and a mel-spectrogram target distribution, CFM directly optimizes the trajectory connecting the two distributions using optimal transport (OT). This facilitates the effective generation of data distributions conditioned on auxiliary information, such as text embedding, audio-visual features, or speaker embeddings, by learning an appropriate conditional vector field~\cite{lipman2022flow, tong2023improving}.

In our framework, data distribution \( p(X) \) is connected to a mel-spectrogram representations, which are connected to a noise distribution \( \pi(X) \) through a continuous trajectory \( \{X_t\}_{t=0}^1 \). Here, \( p(X) \) and \( \pi(X) \) denote the mel-spectrogram data distribution (target distribution) and the noise prior distribution, respectively. The trajectory that continuously transforms \( \pi(X) \) into \( p(X) \), which is optimized by OT-CFM in the sense of optimal transport. Specifically, for \( t \in [0,1] \),
\begin{equation}
    \frac{d}{dt}\phi_t(X)\ = \nu_t^\star(\phi_t(X), t)
\end{equation}
where $\nu_t^\star$ is the optimal vector field that solves the OT problem. During training, we estimate $\nu_t(\phi_t(X), t)$ to approximate $\nu_t^\star$. In our approach, the condition \( \mathbf{c} \) corresponds to audio-visual units with various guidance. Consequently, the evolution of the data is modeled as
\begin{equation}
    \frac{d}{dt}\phi_t(X)\ = \nu_t(\phi_t(X), t \mid \mathbf{c})
\end{equation}
and we train a conditional vector field \( \nu_t \) that integrates both audio and visual features. The OT-CFM optimization objective function is defined by
\begin{equation}
    \min_\theta \, \mathbb{E}_{t,\phi_t(X) \mid \mathbf{c}} \bigl[\|\nu_t(\phi_t(X), t \mid \mathbf{c}) - \nu_t^\star(\phi_t(X), t)\|^2\bigr]
\end{equation}
where \( \nu^\star \) is approximated during training via score matching or stochastic path sampling techniques.
\begin{figure*}[!t]
    \centering
    \begin{subfigure}{0.56\textwidth}
        \centering
        \includegraphics[width=\textwidth]{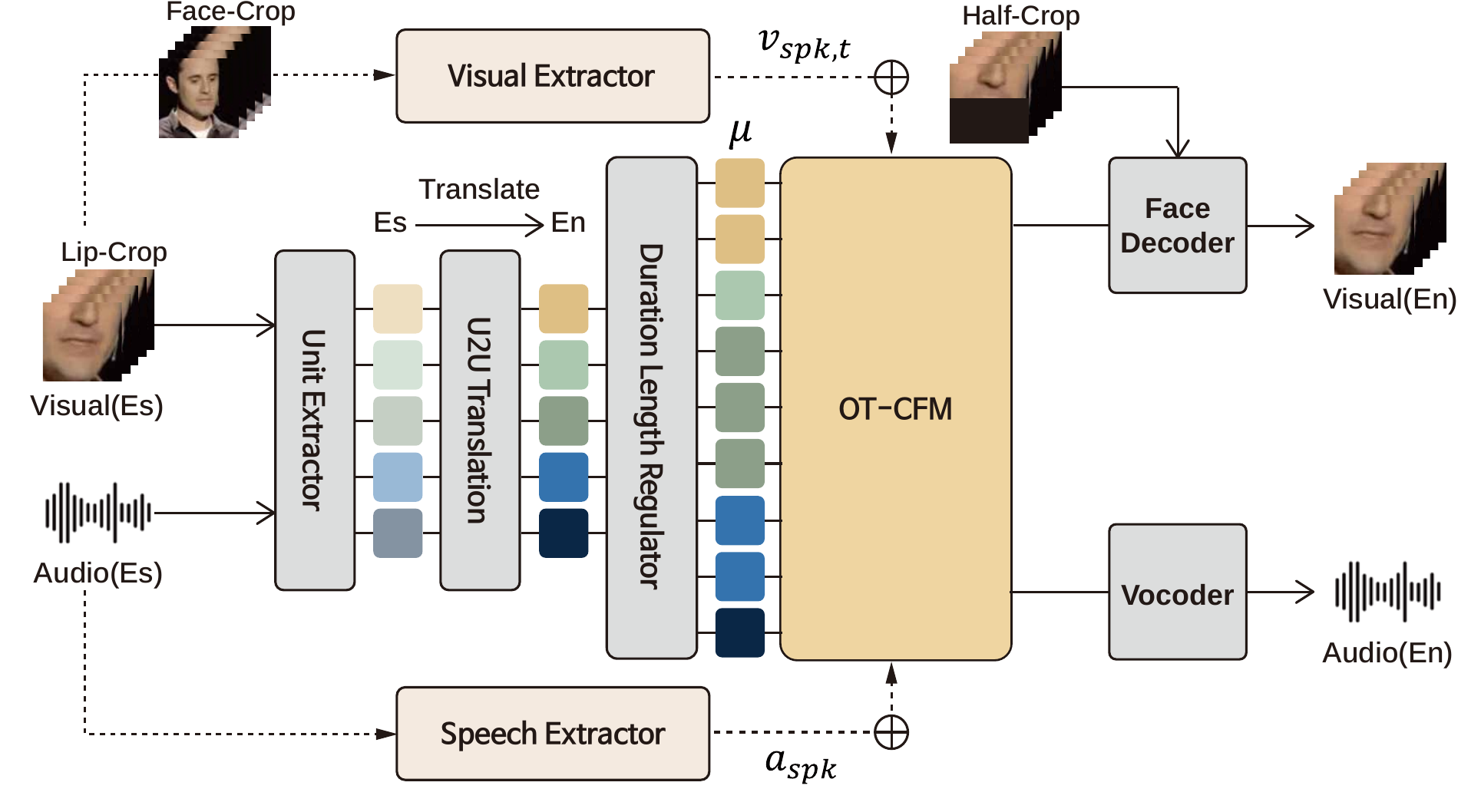}
        \caption{Overall framework of \ourmodel}
        \label{fig:mavflow_b}
    \end{subfigure}
    \hfill
    \begin{subfigure}{0.4\textwidth}
        \centering
        \includegraphics[width=\textwidth]{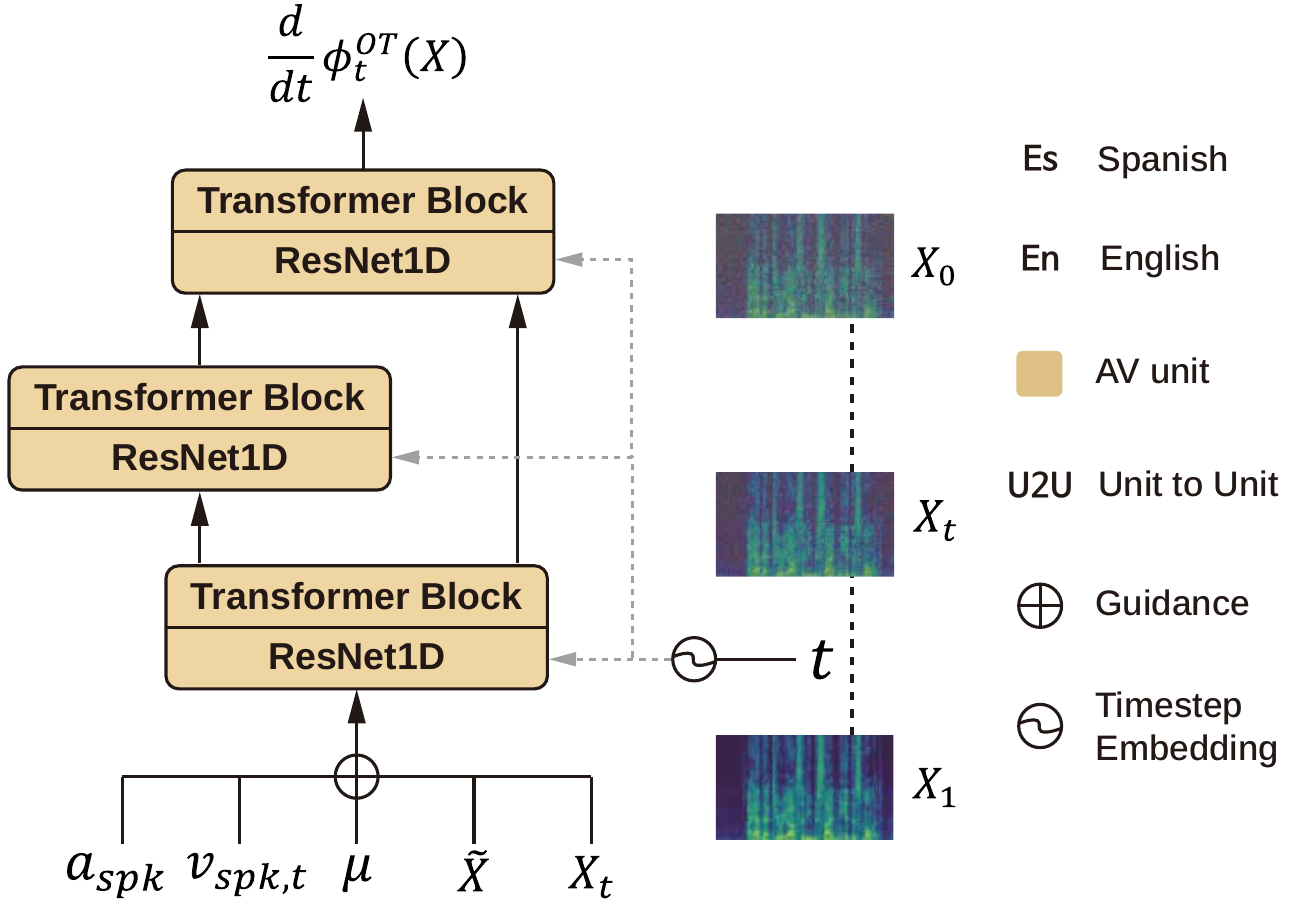}
        \vspace{1mm}
        \caption{OT-CFM decoder structure}
        \label{fig:mavflow_a}
    \end{subfigure}
    \caption{Overall framework and detailed architecture of \ourmodel. (a)\,An overview of the proposed \ourmodel translation system. (b)\,OT-CFM's Transformer decoder structure with multimodal guidance $\mathbf{a}_\mathrm{spk}$ and $\mathbf{v}_\mathrm{spk,t}$ to generate guided mel-spectrogram.} 
    \label{fig:mavflow}
\end{figure*}

\section{\ourmodel}
\label{sec:method}
To effectively preserve speaker-specific characteristics such as voice consistency and facial expressions in multilingual audio-visual translation, we introduce \ourmodel, comprising four main stages: \textit{(i) Audio-Visual Speech Unit Translation}, which we have outlined in Section~\ref{subsec:unit_translation}; \textit{(ii) Duration Length Regulator}; \textit{(iii) Multimodal Guidance}; and \textit{(iv) CFM-based Zero-Shot AV-Renderer} which effectively integrates paralinguistic multimodal guidance with linguistic audio-visual units to synthesize audio-visual outputs. The overall architecture and pipeline of \ourmodel are illustrated in Figure~\ref{fig:mavflow}.

\subsection{Duration Length Regulator}
Since the output of the U2U translation module is deduplicated, it is necessary to predict and expand the duration of each unit. To achieve this, we employ a \textit{Duration Length Regulator}, adapting duration prediction concepts previously explored in TTS synthesis specifically for our audio-visual translation task. We adopt a similar duration prediction structure and loss function from AV2AV~\cite{choi2024av2av}, using two 1D-convolution layers with a classifier, where the objective function is the MSE loss in the log domain. However, our \textit{Duration Length Regulator} differs in that it interpolates the generated audio to match the length of the original source audio. This design addresses a critical constraint in real-world movie dubbing scenarios, where the video length must remain consistent before and after translation—an aspect not considered in AV2AV.

\subsection{Multimodal Guidance}
\label{subsec:guidance}
In the process of generating mel-spectrograms based on the linguistic information of the AV speech unit, conventional methods~\cite{choi2024av2av} relying solely on audio often fall short in accurately capturing visual aspects such as the speaker's emotional state or facial expressions. Particularly in multilingual audio-visual translation, preserving the speaker's natural characteristics requires incorporating not only vocal attributes but also paralinguistic elements like facial expressions. To address this limitation, our approach introduces multimodal guidance by integrating both \emph{speaker voice embeddings} extracted from audio and \emph{speaker facial emotion embeddings} derived from visual inputs. This dual guidance strategy enables a clearer transmission of speaker-specific traits across both modalities, resulting in more consistent and natural synthesis of voice and emotional expression in multilingual translation scenarios. 

\vspace{-4pt}
\paragraph{Speaker voice embedding.}
To capture paralinguistic elements from the audio modality, we use a pretrained speaker encoder to extract x-vectors~\cite{snyder2018x}, which encode the speaker’s unique timbre and speaking style. These robust speaker embeddings are particularly suitable for our cross-lingual scenario. Specifically, in training phase we calculate x-vectors for multiple utterances from the same speaker, then average them to form a \emph{speaker-level} embedding $\mathbf{a}_{\mathrm{spk}}$:
\vspace{-3pt}
\begin{equation}
    \mathbf{a}_{\mathrm{spk}} = \frac{1}{N}\sum_{i=1}^{N} \mathbf{a}_{\mathrm{utt},i}
    \vspace{-1pt}
\end{equation}
where $\mathbf{a}_{\mathrm{utt},i}$ is the x-vector extracted from the $i$-th utterance of a given speaker, and $N$ is the total number of utterances for that speaker. The use of such an averaged speaker embedding allows the model to learn general speaker information during training, thereby enabling the model to robustly learn common speaker traits. 

To guide mel-spectrogram generation using a global speaker representation, we concatenate this speaker embedding with the latent feature of each frame. This frame-level concatenation ensures that synthesized speech consistently reflects the speaker's unique characteristics. During inference, we directly utilize utterance-specific embedding $\mathbf{a}_{\mathrm{utt},i}$, capturing and preserving fine-grained variations unique to each utterance.
By employing distinct speaker embeddings for each phase, our model learns general, global paralinguistic information during training, while effectively capturing local, utterance-specific paralinguistic variations during inference, ultimately enhancing the quality of the generated mel-spectrograms.

\paragraph{Speaker facial emotion embedding.}
In addition to audio cues, we incorporate paralinguistic speaker face emotional embeddings to ensure that the model also learns from visual characteristics of the speaker. We adopt EmoFAN~\cite{toisoul2021estimation} to extract facial emotion embeddings from each frame. Specifically, emotional information in a speaker's utterance can vary dynamically across frames. For instance, a speaker may start smiling partway through an utterance or shift emotional states over time. Thus, distinct emotion embeddings 
$\mathbf{v}_{\mathrm{spk},t}$ are added as guidance for generating each mel-spectrogram frame $X_t$:
\begin{equation}
    \mathbf{v}_{\mathrm{spk}, t} = \text{Emo}(\mathbf{f}_\mathrm{t})
\end{equation}
\noindent where $\mathbf{v}_{\mathrm{spk},t}$ denotes the face embedding of the $t$-th sampled frame $\mathbf{f}_t$, and $\text{Emo}(\cdot)$ is facial emotion extractor. This reflects a distinctive aspect of the cross-lingual scenario, where frame-level speaker audio characteristics vary according to language-specific phonetic and prosodic differences (\eg, variations in accent, intonation patterns, rhythm, and stress placement), whereas emotional information remains consistent across languages. 

\subsection{CFM-based Zero-Shot AV-Renderer}
The CFM-based AV-Renderer integrates translated AV units containing linguistic information with multimodal guidance carrying paralinguistic features. The AV units utilize interpolation to effectively synchronize audio and visual modalities temporally. Additionally, speaker embeddings, as global paralinguistic features, are uniformly added to every frame to maintain consistent emotional and speaker characteristics, while visual embeddings are applied individually across temporal frames. This ensures both temporal and linguistic coherence, resulting in a mel-spectrogram that naturally blends the speaker's facial expressions with their acoustic properties.

\paragraph{Guided mel generation.} To effectively synthesize intermediate mel-spectrograms from audio-visual units, \ourmodel incorporates multimodal information to capture paralinguistic features, as introduced in Section~\ref{subsec:guidance}. Specifically, we employ CFM to guide the mel-spectrogram generation process, optimizing an objective defined as:

\vspace{-0.3cm}
\begin{equation}
\begin{aligned}
\mathcal{L}_{OT-CFM} &=
    \mathbb{E}_{t, p_0(X_0), q(X_1)} \Big[
    \omega_t\bigl(\phi_t^{OT}(X_0, X_1)\mid X_1\bigr) \\
&\quad
    -\,\nu_t \bigl(\phi_t^{OT}(X_0, X_1)\mid \theta\bigl)
    \Big].
\end{aligned}
\end{equation}

\noindent where $\phi_t^{OT} (X_0, X_1)$ is $(1 - (1 - \sigma)t) X_0 + t X_1$ and $\omega_t \left( \phi_t^{OT} (X_0, X_1) \middle| X_1 \right)$ is $X_1 - (1 - \sigma)X_0$. \vspace{2pt}

The multimodal embeddings consist of a global speaker embedding $\mathbf{a}_\mathrm{spk}$, uniformly applied across all frames, and a frame-level emotion embedding $\mathbf{v}_\mathrm{spk, t}$, dynamically varying per timestep. These embeddings, together with the linguistic speech tokens $\{\mu_l\}_{1:L}$ and the masked mel-spectrogram $\tilde{X}_1$, are jointly fed into the neural network $N_\theta$ to match the conditional vector field parameterized by $\theta$, facilitating the integration of global speaker characteristics and local emotional dynamics (as shown in Figure~\ref{fig:mavflow_b}).

\vspace{-0.3cm}
\begin{equation}
\begin{array}{l}
\nu_t \bigl(\phi_t^{OT}(X_0, X_1)\mid\theta\bigl) \\
[0.5em]\hspace{-0.3cm} = N_\theta\bigl(\phi_t^{OT} (X_0, X_1), t; \mathbf{a}_\mathrm{spk}, \mathbf{v}_\mathrm{spk, t}, \{\mu_l\}_{1:L}, \tilde{X}_1 \bigl)
\end{array}
\end{equation}

This strategic utilization of multimodal embeddings, which integrates complementary global speaker identity from audio and frame-level emotional dynamics from visual inputs, plays a crucial role in improving naturalness and speaker consistency in multilingual audio-visual translation.

\section{Experiments}
\label{sec:experiments}

\subsection{Implementation Details}
\paragraph{Dataset.} For training and evaluation, we utilize MuAViC~\cite{anwar2023muavic}, a multilingual audio-visual corpus comprising 1,200 hours of transcribed speech from thousands of speakers, curated from LRS3~\cite{afouras2018lrs3} and mTEDx~\cite{salesky2021multilingual}. We use five languages: English, Spanish, French, Italian, and Portuguese. Since MuAViC does not contain emotion labels, we employ an additional dataset, CREMA-D~\cite{cao2014crema}, for emotion evaluation. CREMA-D consists of 7,442 short video clips featuring 91 adult actors expressing six different emotions: anger, disgust, fear, happy, neutral, and sad. Each clip captures an actor uttering a sentence while simultaneously providing facial expressions and vocal information, making it a suitable dataset for evaluating our model’s performance on emotional maintenance.

\vspace{-10pt}

\paragraph{Model description.}
\ourmodel uses the CFM model pretrained on the LibriTTS~\cite{zen2019libritts} as the initial point for more efficient learning. The model is trained on 8 RTX A6000 GPUs with a constant learning rate of 0.0001. The speaker embedding and emotional embedding extracted from each audio and visual input—originally 192 and 256 dimensions, respectively—are compressed to an 80-dimensional representation and used as guidance for the OT-CFM. To convert the generated mel-spectrogram into a raw audio waveform, we train HiFi-GAN~\cite{kong2020hifi} on the LRS3 dataset. We use the same multi-scale L1 and discriminator loss functions proposed in HiFi-GAN. For precise lip-sync and facial expression generation, we use pretrained Wav2Lip~\cite{prajwal2020lip} on the LRS2~\cite{son2017lip} dataset. Details about the inference time are provided in Appendix~\ref{sec:sup_inference_speed}.

\subsection{Baseline Methods}
There exist only two textless systems, AV2AV~\cite{choi2024av2av} and Transface~\cite{cheng2023transface}, that directly utilize units without generating text in the intermediate process. However, our goal is to develop a zero-shot model that generates translated speech while maximally preserving the original speaker's paralinguistics. Therefore, Transface, which follows a similar approach to AV2AV but does not incorporate additional speaker embeddings—thus not supporting zero-shot audio generation—was excluded from our comparison. Accordingly, for reasonable performance comparison, we establish baselines by combining existing systems in a cascaded manner and compare our proposed method against them. Specifically, the cascaded systems are built based on the latest off-the-shelf pre-trained models such as AVSR~\cite{anwar2023muavic}, ASR~\cite{anwar2023muavic}, AV2T~\cite{anwar2023muavic}, A2T~\cite{anwar2023muavic}, NMT~\cite{casanova2024xtts}, TTS~\cite{ casanova2022yourtts, casanova2024xtts}, and TFG~\cite{prajwal2020lip}.

\subsection{Evaluation}
\vspace{-3pt}
\paragraph{Audio evaluation.} 
We assess our model by using speaker similarity metrics. SS (speaker similarity) leverages ERes2Net~\cite{chen2023enhanced}, providing a robust measure of how closely the synthesized speech matches the target speaker’s identity. ERes2Net is a widely used model trained on the VoxCeleb2~\cite{chung2018voxceleb2} dataset for speaker classification. Since SS alone is insufficient to evaluate the temporal alignment between generated and target mel-spectrograms, we additionally adopt Mel Cepstral Distortion with Dynamic Time Warping (MCD-DTW)~\cite{chen2022v2c} and its speech-length weighted variant (MCD-DTW-SL)~\cite{chen2022v2c}. The SL variant further accounts for speech duration, providing a more comprehensive quality metric. 
We then examine translation quality with the ASR-BLEU score. Specifically, an ASR system is used to transcribe the generated audio, and the resulting text is compared against the ground-truth transcription to calculate the BLEU score~\cite{papineni2002bleu}. Additionally, to evaluate the accuracy of emotion recognition, we assess the audio generated by each system using the pretrained emotion2vec~\cite{ma2023emotion2vec}.

\vspace{-10pt}

\paragraph{Visual evaluation.} For visual quality assessment, we employ Lip Sync Error (LSE) confidence and distance (-C/-D)~\cite{prajwal2020lip} and Fréchet Inception Distance (FID)~\cite{heusel2017gans}, where the LSE metrics quantify the synchronization accuracy of lip movements relative to the audio, while FID measures the distributional similarity between generated frames and real images. Additionally, to measure emotional accuracy from the generated visual frames, we utilize a 6-class\footnote{Neutral, Happy, Sad, Angry, Disgust, and Fear} pretrained MAE-DFER~\cite{sun2023mae} model for emotion classification. Also, emotion embedding cosine similarity (ES) is used to complement class-wise accuracy, which may miss subtle emotional variations due to its fixed set of classes.

\vspace{-10pt}

\begin{table}[!t]
    \centering
    \caption{Comparison of zero-shot speaker similarity scores between X-En translated speech and native speech for traditional cascaded systems and direct textless systems. En: English, Es: Spanish, Fr: French, It: Italian, Pt: Portuguese.}
    \label{tab:ss}
    \vspace{-5pt}
    \resizebox{0.8\linewidth}{!}{%
    \begin{tabular}{l l c c c c}
        \toprule
        & \textbf{Method} 
        & \textbf{SS} \(\uparrow\)
        & \textbf{DTW} \(\downarrow\)
        & \textbf{DTW-SL} \(\downarrow\) \\
        \midrule
        
        \multirow{6}{*}{\rotatebox{90}{(a) Es-En}}
        & GT (Es audio)
          & 1.0 & 0.0 & 0.0 \\
        & 4-Stage Cascaded System\textcolor{red}{$^a$}
          & 0.42 & 11.41 & 17.07 \\
        & 3-Stage Cascaded System\textcolor{red}{$^b$}
          & 0.42 & 11.46 & 16.74 \\
        & 2-Stage Cascaded System\textcolor{red}{$^c$}
          & 0.07 & 11.23 & 14.18 \\
        & Direct System (AV2AV)
          & 0.35 & 9.96 & 12.94 \\
        & \textbf{\ourmodel (ours)} 
          & \textbf{0.49} & \textbf{9.60} & \textbf{12.47} \\
        \midrule
        
        \multirow{6}{*}{\rotatebox{90}{(b) Fr-En}}
        & GT (Fr audio)
          & 1.0 & 0.0 & 0.0 \\
        & 4-Stage Cascaded System 
          & 0.34 & 10.75 & 17.00 \\
        & 3-Stage Cascaded System 
          & 0.35 & 10.90 & 16.97 \\
        & 2-Stage Cascaded System 
          & 0.02 & 10.78 & 13.79 \\
        & Direct System (AV2AV) 
          & 0.31 & 9.92 & 12.46 \\
        & \textbf{\ourmodel (ours)} 
          & \textbf{0.51} & \textbf{8.76} & \textbf{10.97} \\
        \midrule
        
        \multirow{6}{*}{\rotatebox{90}{(c) It-En}}
        & GT (It audio)
          & 1.0 & 0.0 & 0.0 \\
        & 4-Stage Cascaded System
          & 0.41 & 11.91 & 17.27 \\
        & 3-Stage Cascaded System
          & 0.41 & 11.80 & 16.79 \\
        & 2-Stage Cascaded System 
          & 0.05 & 11.23 & 13.70 \\
        & Direct System (AV2AV) 
          & 0.37 & 10.44 & 14.75 \\
        & \textbf{\ourmodel (ours)} 
          & \textbf{0.53} & \textbf{9.36} & \textbf{11.43} \\
        \midrule
        
        \multirow{6}{*}{\rotatebox{90}{(d) Pt-En}}
        & GT (Pt audio)
          & 1.0 & 0.0 & 0.0 \\
        & 4-Stage Cascaded System 
          & 0.36 & 11.12 & 17.89 \\
        & 3-Stage Cascaded System 
          & 0.35 & 10.97 & 17.65 \\
        & 2-Stage Cascaded System 
          & 0.11 & 10.89 & 13.72 \\
        & Direct System (AV2AV) 
          & 0.30 & 9.82 & 12.38 \\
        & \textbf{\ourmodel (ours)} 
          & \textbf{0.48} & \textbf{9.14} & \textbf{11.53} \\
        \bottomrule
    \end{tabular}
    }
    \begin{flushleft}
    \vspace{-5pt}
    \scriptsize{\textcolor{red}{$^a$}AVSR\,~\cite{anwar2023muavic}\,+\,NMT\,~\cite{fan2021beyond}\,+\,TTS\,~\cite{casanova2024xtts}\,+\,TFG~\cite{prajwal2020lip} \\ \textcolor{red}{$^b$}AV2T\,~\cite{anwar2023muavic}\,+\,TTS\,~\cite{casanova2024xtts}\,+\,TFG~\cite{prajwal2020lip} \\ \textcolor{red}{$^c$}A2A\,~\cite{kim2024textless}\,+\,TFG~\cite{prajwal2020lip}}
    \vspace{-10pt}
    \end{flushleft}
    \vspace{-4pt}
\end{table}

\paragraph{Human evaluation.} We have conducted subjective evaluations to capture the human perception of generated audio quality. We perform a Mean Opinion Score (MOS) test that includes two factors: MOS-Similarity, to gauge how closely the synthesized speech resembles the target speaker’s voice, and MOS-Naturalness, which evaluates fluency and overall realism. We have recruited 21 participants, each rating a total of 8 audio samples per method. Our evaluation set consists of four different methods: \ourmodel, a 4-stage cascaded system, a 3-stage cascaded system, and AV2AV. To maintain objectivity and avoid excessive evaluations by the assessors, the 2-stage cascaded system, which showed relatively poor performance in Table~\ref{tab:ss}, was excluded. Additionally, since the ground truth audio is in the original language before translation, it was excluded to ensure fairness in the evaluation.

\begin{table}[!t]
    \centering
    \caption{Comparison of zero-shot speaker similarity scores of generated audio for traditional cascaded systems and direct systems, with additional emotion evaluation on the CREMA-D dataset.}
    \label{tab:emotion}
    \vspace{-5pt}
    \resizebox{0.9\linewidth}{!}{%
    \begin{tabular}{l c c c c}
        \toprule 
        \textbf{Method} & \!\!\textbf{Emo-Acc (\%)} \(\uparrow\)\!\! & \textbf{SS} \(\uparrow\)  & \textbf{DTW} \(\downarrow\) & \textbf{DTW-SL} \(\downarrow\) \\
        \midrule
        GT & 81.95 & 1.0 & 0.0 & 0.0  \\
        GT Mel + Vocoder & 68.41 & 0.76 & 1.75 & 1.75 \\
        \midrule
        ASR + YourTTS\,\cite{casanova2022yourtts} & 17.52 & 0.40 & 9.02 & 11.78\\
        ASR + XTTS\,\cite{casanova2024xtts} & 28.55 & \textbf{0.46} & 11.98 & 17.68\\
        Direct System (AV2AV) & 33.66 & 0.33 & 7.84 & 7.88 \\
        \textbf{\ourmodel (ours)} & \textbf{36.46} & 0.39 & \textbf{7.30} & \textbf{7.36} \\
        \bottomrule
    \end{tabular}
    }
    \vspace{-4pt}
\end{table}

\subsection{Zero-shot Audio Translation Result}
\paragraph{Speaker voice similarity.} In Table~\ref{tab:ss}, we evaluate the speaker similarity between the original speech and the speech generated after translation by our model and baseline models. As a result, \ourmodel generates the translated audio that has the highest speaker similarity score with the original voice, compared to the cascaded system and the baseline direct system (AV2AV). This implies that our audio-visual guidance demonstrates outstanding performance in preserving the speaker's identity. In addition, \ourmodel demonstrates superior performance relative to the baseline on the MCD-DTW and MCD-DTW-SL metrics, confirming that the speaker’s pronunciation and timbre are well maintained. In particular, since MCD-DTW-SL also reflects duration consistency, this indicates that our duration length regulator has been effective. These results were obtained using speech generated by translating four source languages—Spanish, French, Italian, and Portuguese—into English. In generating the final translated speech, the speaker embedding extracted from the non-translated original speech and the emotion embedding extracted from the face were used as guidance for the renderer.

\vspace{-10pt}

\paragraph{Emotion evaluation.} To evaluate how accurately the emotion in the speech generated after translation reflects the emotion of the original speech, we compare the proposed model with the baseline model (AV2AV) using the CREMA-D dataset. The evaluation is based on the emotional accuracy calculated by the emo2vec model, which examines how the target speech (the synthesized speech after translation) is classified into ground-truth emotion categories.
In Table~\ref{tab:emotion}, the emotion2vec~\cite{ma2023emotion2vec} model achieves approximately 82\% classification accuracy on the ground-truth(GT) audio, serving as an upper bound for the emotion recognition model itself. In this experiment, our model achieves 36.5\% emotional accuracy (+2.8\%, +7.91\%, and +18.94\% compared to AV2AV, ASR\,+\,YourTTS~\cite{casanova2022yourtts}, and ASR\,+\,XTTS~\cite{casanova2024xtts} respectively), suggesting that it successfully synthesizes speech that preserves emotional traits.
\vspace{-10pt}
\begin{table}[!t]
    \centering
    \small
    \caption{Translation quality (ASR-BLEU score) for X-En translation comparison with cascaded system.}
    \label{tab:asr-bleu}
    \vspace{-5pt}
    \addtolength{\tabcolsep}{-1.5pt}
    \renewcommand{\arraystretch}{1.05}
    \resizebox{\linewidth}{!}{
    \begin{tabular}{llcccc}
        \toprule
         & \textbf{Translation} & \multicolumn{4}{c}{\textbf{X-En}} \\
        \cline{3-6}
        \textbf{Method} & \textbf{Modality} 
        & Es-En & Fr-En & It-En & Pt-En \\
        \midrule
        \multicolumn{2}{l}{~$\bullet$~\textit{\textbf{4-Stage}}} \\
        ASR + NMT + TTS + TFG & A$\rightarrow$AV & 28.66 & 30.55 & 23.54 & 26.14 \\
        AVSR + NMT + TTS + TFG & AV$\rightarrow$AV & \textbf{28.70} & 29.21 & \textbf{24.54} & \textbf{26.30} \\
        
        \multicolumn{2}{l}{~$\bullet$~\textit{\textbf{3-Stage}}} \\
        A2T + TTS + TFG & A$\rightarrow$AV & 24.06 & 27.01 & 21.92 & 24.11 \\
        AV2T + TTS + TFG & AV$\rightarrow$AV & 24.61 & 26.90 & 22.33 & 24.83 \\
        
        \multicolumn{2}{l}{~$\bullet$~\textit{\textbf{2-Stage (Textless)}}} \\
        A2A + TFG & A$\rightarrow$AV & 26.15 & 30.14 & 22.41 & 23.77 \\
        \midrule
        
        \multicolumn{2}{l}{~$\bullet$~\textit{\textbf{Direct (Textless)}}} \\
        AV2AV & AV$\rightarrow$AV & 26.57 & 31.27 & 23.24 & 24.51 \\
        \textbf{\ourmodel (ours)} & AV$\rightarrow$AV & 26.97 & \textbf{31.33} & 23.43 & 24.97 \\
        
        \bottomrule
    \end{tabular}
    }
\end{table}

\paragraph{Translation quality.} In Table~\ref{tab:asr-bleu}, we evaluate the translation quality using the ASR-BLEU score for different language pairs. The result demonstrates that \ourmodel achieves improved translation performance compared to AV2AV. Since we generated speech using the same unit translation model as AV2AV, this confirms that our model produces more accurate speech outputs when given identical units. These results suggest that our model leverages the structural advantages of CFM to enhance feature matching and rendering, thereby increasing both the accuracy and consistency of the generated speech. Furthermore, \ourmodel exhibits competitive translation quality when compared to the cascaded systems. This result implies that our dual modality guidance does not impair semantic quality during translation, which is also critical in AV2AV applications, while better preserving paralinguistic elements (as seen in Tables~\ref{tab:ss}--\ref{tab:emotion}).

\vspace{-8pt}
\begin{table}[!t]
    \centering
    \caption{Comparison of MOS scores between X-En translated speech and native speech for traditional cascaded systems and direct systems.}
    \label{tab:mos}
    \vspace{-5pt}
    \renewcommand{\arraystretch}{1.1}
    \resizebox{0.85\linewidth}{!}{
    \begin{tabular}{l c c}
        \toprule
        \textbf{Method} & \textbf{Similarity} $\uparrow$ & \textbf{Naturalness} $\uparrow$ \\
        \midrule
        4-Stage Cascaded System & 2.81 & 3.29 \\
        3-Stage Cascaded System & 2.89 & 3.25 \\
        Direct System (AV2AV) & 3.33 & 3.58 \\
        \textbf{\ourmodel (ours)} & \textbf{3.49} & \textbf{4.01} \\
        \bottomrule
    \end{tabular}
}
\vspace{-4pt}
\end{table}

\paragraph{Subjective evaluation.} To evaluate the naturalness of the generated speech, we assessed the MOS scores for the translated speech from the MuAViC dataset generated by each system in Table~\ref{tab:mos}. The evaluation results show that our naturalness quality achieved higher MOS scores (3.49 for Similarity, 4.01 for Naturalness) compared to other cascade systems and AV2AV (3.33 for Similarity, 3.58 for Naturalness).

\begin{table}[!t]
    \centering
    \small
    \caption{Reconstruction visual quality performance on LRS3.}
    \label{tab:visual}
    \vspace{-5pt}
    \addtolength{\tabcolsep}{-1pt}
    \renewcommand{\arraystretch}{1.2}
    \resizebox{0.82\linewidth}{!}{
    \begin{tabular}{c l c c c c}
        \toprule
        \textbf{ID} & \textbf{Method} & \textbf{LSE-C} $\uparrow$ & \textbf{LSE-D} $\downarrow$ & \textbf{FID} $\downarrow$ \\
        \midrule
        \multicolumn{5}{l}{~$\bullet$~\textbf{\textit{Ground Truth}}} \\
        C1 & GT Audio-Visual & 7.63 & 6.89 & - \\
        \hline
        \multicolumn{5}{l}{~$\bullet$~\textbf{\textit{Cascaded System}}} \\
        C2 & GT Audio + TFG & 8.23 & \textbf{6.75} & 5.66 \\
        C3 & GT Text + TTS + TFG & 7.01 & 7.49 & \textbf{5.38} \\
        \hline
        \multicolumn{5}{l}{~$\bullet$~\textbf{\textit{AV2AV}}} \\
        C4 & GT AV Speech Unit & 7.43 & 7.30 & 6.30 \\
        \hline
        \multicolumn{5}{l}{~$\bullet$~\textbf{\textit{\ourmodel (ours)}}} \\ 
        C5 & GT AV Speech Unit & \textbf{8.30} & 6.81 & 5.69 \\  
        \bottomrule
    \end{tabular}
    }
    \vspace{-4pt}
\end{table}

\begin{figure*}[!t]
    \centering
    \begin{minipage}[t]{0.65\textwidth}
        \centering
        \vspace{-170pt}
        \includegraphics[width=\textwidth]{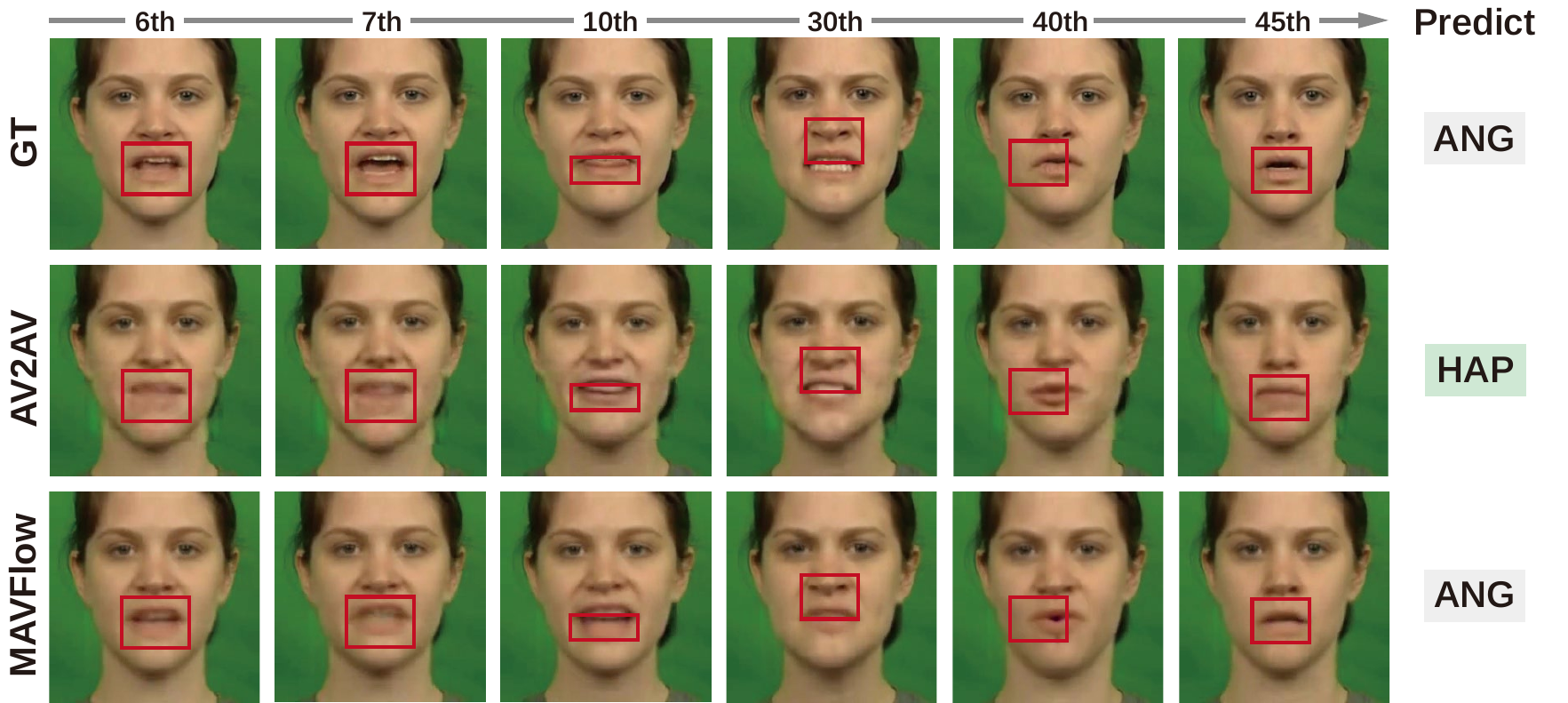}
        \captionof{figure}{Visual analysis of emotional representation in generated videos. By applying speaker facial emotion embeddings at each frame, our approach enhances frame-level emotional accuracy. As highlighted by the rectangular boxes, our method effectively resolves emotion misclassification issues found in the AV2AV.}
        \label{fig:visualization}
    \end{minipage}
    \hfill
    \begin{minipage}[b]{0.3\textwidth}
        \centering
        \small
        \captionof{table}{Visual emotion recognition accuracy (Emo-Acc) and emotion embedding cosine similarity (ES) measured from the generated visual results on CREMA-D.}
        \label{tab:visual_emo}
        \vspace{-5pt}
        \resizebox{\linewidth}{!}{%
            \begin{tabular}{l c c}
                \toprule 
                \textbf{Method} & \textbf{Emo-Acc (\%)} $\uparrow$ & \textbf{ES} $\uparrow$\\
                \midrule
                GT & 76.83 & 1.00\\
                AV2AV & 67.20 & 0.87\\
                \textbf{\ourmodel (ours)} & \textbf{72.68} & \textbf{0.92} \\
                \bottomrule
            \end{tabular}
        }
        
        \captionof{table}{Audio emotion accuracy and embedding cosine similarity (ES) after additional training (+: additional training on CREMA-D).}
        \label{tab:add_training}
        \resizebox{\linewidth}{!}{%
            \begin{tabular}{l c c c}
                \toprule 
                \textbf{Method} & \textbf{Emo-Acc (\%)} $\uparrow$ & \textbf{ES} $\uparrow$ & \textbf{SS} $\uparrow$\\
                \midrule
                {AV2AV} & 33.66 & 0.84 & 0.33  \\
                \textbf{\ourmodel} & 36.46 & 0.86 & 0.39 \\
                \textbf{\ourmodel+} & \textbf{51.46} & \textbf{0.90} & \textbf{0.49} \\
                \bottomrule
            \end{tabular}
        }
    \end{minipage}
\end{figure*}

\subsection{Zero-shot Video Translation Result}
\paragraph{Visual generation quality.} In Table~\ref{tab:visual}, we evaluated the visual quality of the generated videos and the synchronization between the audio and visual components. \ourmodel achieves an LSE-C score of 8.30, outperforming all baseline methods. Particularly, when compared to AV2AV (C4), which has a similar direct synchronization structure to ours, \ourmodel demonstrates significant improvements across all metrics: LSE-C ($+$0.87), LSE-D ($-$0.49), and FID ($-$0.61). These results indicate that audio-visual guidance not only enhances the consistency of synthesized speech but also positively affects face generation quality. 

Specifically, the high LSE-C score highlights a strong correlation between the generated audio and video, suggesting that \ourmodel effectively utilized visual embeddings. In other words, our model successfully integrated latent visual information from the initial stages of mel-spectrogram generation through visual guidance. Additionally, the synthesized face images, based on high-quality mel-spectrograms, also exhibited competitive performance in the FID metric, confirming the generation of more natural and realistic faces.

\vspace{-10pt}

\paragraph{Visual emotional quality.} In Figure~\ref{fig:visualization}, we analyze the generated visual quality on the CREMA-D dataset and evaluate whether each generated visual frame accurately reflects the speaker's emotion using a visual emotion recognition model. Through this evaluation, we confirm that our proposed method, which applies visual embedding at the frame level, effectively captures the original emotional state of the speaker over time. For instance, in Figure~\ref{fig:visualization}, the AV2AV method incorrectly predicted `HAP' (Happy) for an original video labeled with `ANG' (Anger). Upon examining each frame closely, it becomes clear that the video generated by AV2AV fails to adequately express the anger emotion, particularly around the mouth, compared to the ground truth video. These visual observations are further supported by the quantitative results in Table~\ref{tab:visual_emo}, which show that while AV2AV's visual emotion recognition performance decreases compared to the ground truth, our proposed method demonstrates better preservation not only in terms of accuracy (Emo-Acc) but also in embedding similarity (ES). Additional visual quality can be referred to in Appendix~\ref{sec:supp_visual}.

\subsection{Ablation Study}
\paragraph{Additional training on emotion dataset.} In Table~\ref{tab:emotion}, we only trained our model on the MuAViC dataset to enable zero-shot evaluation on the unseen CREMA-D benchmark. However, additional training on emotion-rich audio-visual datasets can significantly enhance emotion transfer performance. In Table~\ref{tab:add_training}, we lightly uptrained MAVFlow on CREMA-D training datasets (referred to as MAVFlow+), resulting in a notable increase in Emo-Acc from 36.46\% to 51.46\% as well as an improvement in ES from 0.86 to 0.90. Since \ourmodel outperformed the baselines using only MuAViC, we expect that incorporating such emotion-rich data would further widen this performance gap.

\paragraph{Effect of audio-visual guidance.} We conducted an ablation study to examine the effect of each modality guidance on audio generation. Table~\ref{tab:ablation} presents the results of evaluating the effect of audio and visual modality guidance on emotion recognition using the CREMA-D audio dataset. Additionally, it includes the analysis of results from translating audio in Es, Fr, It, and Pt to En using the MuAViC dataset, based on the settings outlined in Table~\ref{tab:ss}. The SS and MCD-DTW values in Table~\ref{tab:ablation} were averaged across each language for analysis. As a result, we observed that when both audio and visual guidance were provided, speaker similarity and emotional accuracy improved. One interesting observation is that when visual guidance is provided alone, speaker similarity slightly increases or is maintained (as seen in Table~\ref{tab:ablation}), but Emo-Acc decreases. This suggests that visual guidance alone has a minimal effect on maintaining emotion, and its complementary effect is maximized when combined with audio guidance.
\begin{table}[!t]
    \centering
    \small
    \caption{Ablation study for the effect of modality guidance on CREMA-D and MuAViC translation.}
    \label{tab:ablation}
    \vspace{-5pt}
    \resizebox{0.9\linewidth}{!}{
    \begin{tabular}{c c | c c | c c}
    \toprule
    & & \multicolumn{2}{c|}{CREMA-D} & \multicolumn{2}{c}{MuAViC} \\
    \cmidrule(l{5pt}r{5pt}){3-4} \cmidrule(l{5pt}r{5pt}){5-6}
    \textbf{Audio} & \textbf{Visual} & \textbf{SS} $\uparrow$ & \textbf{Emo-Acc} $\uparrow$ & \textbf{SS} $\uparrow$ & \textbf{DTW} $\downarrow$ \\
    \midrule
    \xmark & \xmark & 0.167 & 28.66 & 0.057 & 10.13 \\
    \xmark & \cmark & 0.174 & 26.83 & 0.056 & 10.73 \\
    \cmark & \xmark & \textbf{0.391} & 35.85 & 0.487 & 7.50 \\
    \cmark & \cmark & 0.388 & \textbf{36.46} & \textbf{0.504} & \textbf{7.37} \\
    \bottomrule
    \end{tabular}
    }
    \vspace{-6pt}
\end{table}

\section{Conclusion}
\label{sec:conclusion}
In this paper, we introduced \ourmodel, a zero-shot audio-visual translation framework utilizing Conditional Flow Matching (CFM) to address speaker consistency challenges inherent in existing AV2AV methods. By effectively integrating paralinguistic characteristics from both audio and visual modalities, \ourmodel significantly enhances speaker consistency across languages without intermediate text representations. Our method leverages discrete speech units and dual-modal guidance to synthesize high-quality mel-spectrograms, resulting in improved lip synchronization, emotional accuracy, and overall visual quality. Experimental evaluations on the MuAViC and CREMA-D datasets confirm that \ourmodel outperforms prior AV2AV methods, establishing it as a robust and efficient solution for multilingual audio-visual translation.
\section*{Acknowledgements}
\label{sec:acknowledgements}
This work was supported by Institute of Information \& communications Technology Planning \& Evaluation (IITP) grant funded by the Korea government (MSIT) [No. 2022-0-00641, XVoice: Multi-Modal Voice Meta Learning], [No. RS-2024-00457882, AI Research Hub Project], and [No. 2019-0-00075, Artificial Intelligence Graduate School Program (KAIST)].
{
    \small
    \bibliographystyle{ieeenat_fullname}
    \bibliography{main}
}
\newpage
\appendix
\begin{center}
    \textsc{\Large Appendix} \\
\end{center}
\section{Qualitative Results and Analysis}
\label{sec:supp_visual}

\begin{figure*}[bt!]
    \centering
    \includegraphics[width=\linewidth]{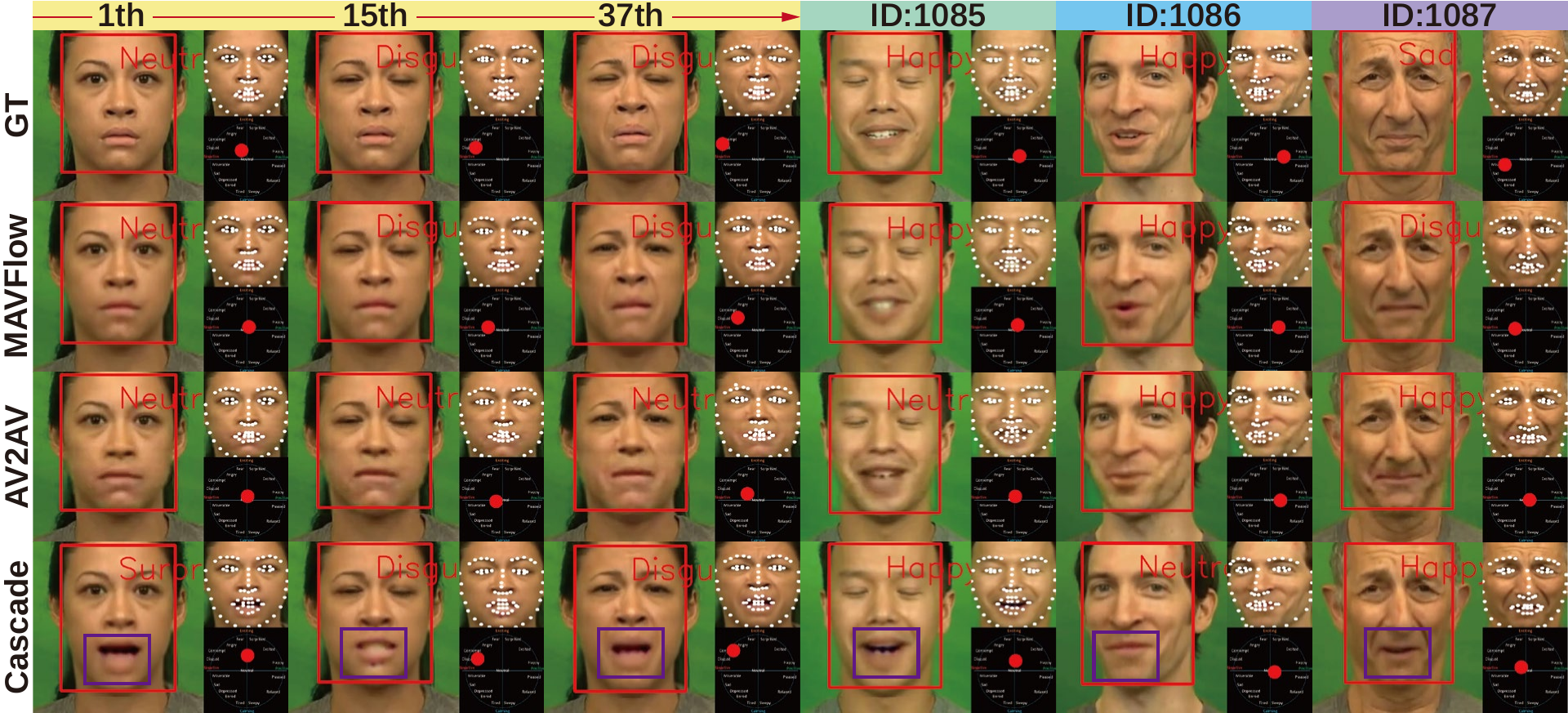}
    \caption{Additional qualitative comparison for frame-level analysis. Each row shows GT, \ourmodel, AV2AV, and Cascade (ASR+XTTS+TFG), respectively.}
    \label{fig:visual}
\end{figure*}

In Figure~\ref{fig:visual}, we present dynamic emotional changes across frames within a single video at the first three frames from neutral to disgust. While \ourmodel effectively captures the emotional change of Ground Truth (GT) video from 15th frame, reflecting the shift starting from the 14th frame. AV2AV fails to reflect the emotion until around the 36th frame. Additionally, overall, \ourmodel better expresses emotions as well as the arousal level, which is indicated by the distance of a red dot from the center. Cascaded systems have been excluded from the comparison due to poor temporal alignment and their inability to embed emotional cues into the audio, which results in TFG output cannot reflect emotional expressions in the video. The DTW and DTW-SL metrics in Table~\ref{tab:ss} and Table~\ref{tab:emotion}, further confirm the notably poor temporal alignment of the cascaded systems.

\section{Class-Wise Emotional Analysis}
\subsection{Audio Emotional Results}
\label{sec:sup_finetuning}
In Table~\ref{tab:emotion_supp}, we evaluate the class-wise emotion recognition accuracy of the generated audio using the pretrained emotion2vec~\cite{ma2023emotion2vec}. Compared to AV2AV, \ourmodel shows slightly lower performance for the Sad, Disgust, and Fear classes, while demonstrating comparable or superior results for Happy, Neutral, and Angry. Notably, \ourmodel exhibits a significant advantage in the Angry class, ultimately achieving better overall performance than AV2AV in both Emo-Acc and ES metrics (as shown in Table~\ref{tab:add_training}). Furthermore, the \ourmodel+ model, trained with additional emotional datasets, achieves improved performance across most emotion classes, with a substantial gain in overall Emo-Acc.
\begin{table}[!ht]
    \centering
    \caption{Class-wise emotion accuracy (\%) of generated audio (+: additional training on CREMA-D).}
    \vspace{-5pt}
    \addtolength{\tabcolsep}{-1.5pt}
    \resizebox{\linewidth}{!}{%
    \begin{tabular}{l c c c c c c c}
        \toprule 
        \textbf{Method} & \!\!\textbf{Happy}\!\! & \textbf{Sad} & \!\!\!\!\textbf{Neutral}\!\!\!\! & \textbf{Angry} & \!\!\textbf{Disgust}\!\! & \textbf{Fear} &  \!\!\!\!\textbf{Emo-Acc}\,\(\uparrow\)\!\!\!\!\\
        \midrule
        {GT} & 89.29 & 85.00 & 89.17 & 89.29 & 77.86 & 62.14 & 81.95 \\
        \hline
        {AV2AV} & 30.00 & \textbf{22.86} & \textbf{80.00} & 28.57 & 30.71 & 16.43 & 33.66  \\
        \textbf{\ourmodel} & 36.43 & 11.43 & \textbf{80.00} & 62.86 & 20.00 & 14.29 & 36.46 \\
        \textbf{\ourmodel+} & \textbf{69.29} & \textbf{22.86} & 66.67 & \textbf{80.71} & \textbf{32.86} & \textbf{38.57} & \textbf{51.46} \\
        \bottomrule
    \end{tabular}
    }
    \label{tab:emotion_supp}
\end{table}

\subsection{Visual Emotional Results}
\label{sec:supp_visual_emotion}
In Table~\ref{tab:emotion2}, we evaluated class-wise visual emotion accuracy using pretrained MAE-DFER~\cite{sun2023mae}. Also, follow MAE-DFER, we report both Unweighted Average Recall (UAR) and Weighted Average Recall (WAR) as evaluation metrics. UAR calculates the average recall by treating each class equally, which helps account for class imbalance, while WAR weights the recall by the number of samples per class, reflecting the actual class distribution in the dataset. As a result, \ourmodel achieved strong performance in terms of both UAR and WAR, particularly excelling in the angry, disgust, and fear emotion classes.

\begin{table*}[!b]
    \centering
    \caption{Class-wise emotion accuracy, unweighted and weighted average recall (UAR\%, WAR\%), and ES of the generated visuals, all measured with MAE-DFER (+: additional training on CREMA-D).}
    \addtolength{\tabcolsep}{-1pt}
    \resizebox{0.7\linewidth}{!}{%
    \begin{tabular}{l c c c c c c c c c}
        \toprule 
        \textbf{Method} & \!\!\textbf{Happy} & \textbf{Sad} & \textbf{Neutral} & \textbf{Angry} & \textbf{Disgust} & \textbf{Fear} & \textbf{UAR} & \textbf{WAR} & \textbf{ES}\\
        \midrule
        {GT} & 97.14 & 67.86 & 76.67 & 78.57 & 87.86 & 52.86 & 76.83 & 76.83 & 1.00 \\
        \midrule
        {ASR+YourTTS+TFG} & 89.86 & 60.71 & 72.88 & 50.71 & 83.57 & 40.00 & 66.29 & 66.05 & 0.85 \\
        {ASR+XTTS+TFG} & 94.93 & 55.00 & 72.03 & 72.86 & 85.71 & 31.43 & 68.66 & 68.50 & 0.91 \\
        {AV2AV} & \textbf{95.00} & \textbf{64.29} & \textbf{79.17} & 62.14 & 77.14 & 27.14 & 67.48 & 67.20 & 0.87 \\
        \textbf{\ourmodel} & \textbf{95.00} & 53.57 & 75.00 & \textbf{80.71} & \textbf{88.57} & \textbf{43.57} & 72.74 & 72.68 & 0.92 \\
        \textbf{\ourmodel+} & \textbf{95.00} & 63.57 & 78.33 & 76.43 & 87.14 & 37.86 & \textbf{73.06} & \textbf{72.93} & \textbf{0.93}\\
        \bottomrule
    \end{tabular}
    }
    \label{tab:emotion2}
\end{table*}

\section{Inference Time Comparison}
\label{sec:sup_inference_speed}
\ourmodel does not rely on intermediate text representations, resulting in faster inference compared to the cascaded system. Furthermore, it is more efficient by applying the speed-friendly CFM module compare to diffusion model. We compared the inference speed using one A6000 GPU, observing processing times of 1.66s for MAVFlow, 1.22s for AV2AV, and 1.75s for the 4-cascaded model to handle a 2.35s audio-visual input through the complete pipeline.

\section{Limitation}
\label{sec:supp_limitation}
\ourmodel currently leverages emotional embeddings only from face and speaker embeddings from audio. However, we believe that incorporating emotional cues from audio (\eg, prosody, timbre, and other paralinguistic features) into the guidance of CFM could further enhance performance. Furthermore, since we directly adopt the unit extractor and unit-to-unit translation modules from previous work~\cite{choi2024av2av}, improving semantic translation quality remains an open challenge.

\end{document}